\documentstyle[12pt,titlepage,epsf]{article}
\newlength{\extraspace}
\newlength{\extraspaces}
\catcode`\@=11
\def\numberbysection{\@addtoreset{equation}{section}
  \def\theequation{\arabic{section}.\arabic{equation}}}
\def\rt{\sqrt}

\def\rmL{{\rm L}}
\def\rmR{{\rm R}}

\def\rmW{{\rm W}}

\def\rmc{{\rm c}}

\def\rmt{{\rm t}}
\def\rmB{{\rm B}}

\def\Pl{{\rm Pl}}

\def\rmHc{{\rm H. c.}}

\def\rmSU{{\rm SU}}
\def\rmU{{\rm U}}

\def\beq{\begin{equation}}
\def\eeq{\end{equation}}
\def\ba{\left(\!\!\begin{array}}
\def\ea{\end{array}\!\!\right)}
\def\bea{\begin{eqnarray}}
\def\eea{\end{eqnarray}}
\def\beas{\begin{eqnarray*}}
\def\eeas{\end{eqnarray*}}

\def\ben{\begin{enumerate}}
\def\een{\end{enumerate}}
\def\bi{\begin{itemize}}
\def\ei{\end{itemize}}
\def\bd{\begin{description}}
\def\ed{\end{description}}
\def\bc{\begin{center}}
\def\ec{\end{center}}

\def\al{\alpha}

\def\G{\Gamma}

\def\e{\epsilon}

\def\th{\theta}

\def\m{\mu}

\def\p{\pi}

\def\s{\sigma}

\def\ta{\tau}

\def\f{\phi}

\def\**{\times}
\def\(({\left(}
\def\)){\right)}
\def\+-{\pm}
\def\-+{\mp}

\begin{document}
\begin{titlepage}
\addtolength{\baselineskip}{.7mm}
\thispagestyle{empty}
\begin{flushright}
TIT/HEP--236 \\
October, 1993
\end{flushright}
\begin{center}
{\large{\bf
Cosmological Baryon Asymmetry in Supersymmetric \\
 Standard Models and Heavy Particle Effects}} \\[15mm]
{\sc Tomoyuki Inui}
\footnote{{\tt JSPS Fellow, e-mail: kankun@phys.titech.ac.jp}}
{,\ \ \sc Tomoyasu Ichihara}
\footnote{\tt e-mail: tomo@phys.titech.ac.jp} \\[3mm]
{\sc Yukihiro Mimura}
\footnote{\tt e-mail: mim@phys.titech.ac.jp}
\ \ and \ \
{\sc Norisuke Sakai}
\footnote{\tt e-mail: nsakai@phys.titech.ac.jp} \\[4mm]
{\it Department of Physics,
Tokyo Institute of Technology, \\[2mm]
Oh-okayama, Meguro, Tokyo 152, Japan} \\
\vfill
{\bf Abstract}\\[5mm]
{\parbox{13cm}{\hspace{5mm}
Cosmological baryon asymmetry $B$ is studied in supersymmetric standard
models, assuming the electroweak reprocessing of $B$ and $L$.
Only when the soft supersymmetry breaking is taken into account,
$B$ is proportional to the primordial $B-L$ in the supersymmetric standard
models. The ratio $B/(B-L)$ is found to be about one percent less than the
nonsupersymmetric case.
Even if the primordial $B-L$ vanishes,
scalar-leptons can be more efficient than leptons to generate
$B$ provided that mixing angles $\th$ among scalar leptons
satisfy $|\th| < 10^{-8} (T/\mbox{GeV})^{\frac12}$.
}}
\end{center}
\vspace*{5ex}
\end{titlepage}
\setcounter{section}{0}
\setcounter{equation}{0}
%
%%%%%%%  Introduction  %%%%%%%%%%%%%%%%%%%%%%%%%%%%%%%%%%%%%%%%%%%
Explaining the cosmological baryon asymmetry in the
evolutionary history of the universe is one of the most fascinating
problems in cosmology and particle theory \cite{Sakharov67}.
%It is now well-established
The elecroweak $B+L$ anomaly interactions
($B$ for baryon number density, $L$ for lepton number density)
provide an efficient mechanism to transmute between $B$ and
$L$ \cite{tHooft76PRLPR}--\cite{KuzminRSM}.
Since
%these processes
they are rapid enough to maintain thermal equilibrium
at least well above the temperature for the electroweak phase
transition, they may wash away (or at least alter) the primordial
baryon asymmetry which has been generated in the early universe.
It has been difficult to produce sufficient baryon asymmetry at the
electroweak phase transition.
Since $B-L$ suffers no anomaly, it is conserved in this electroweak
reprocessing of baryon asymmetry.
We assume that $B$ is obtained from the reprocessing of the primordial $B-L$.
A primordial lepton number $L$ has been proposed as
a source of the baryon asymmetry through the electroweak
reprocessing \cite{FukugitaY}.
The resulting ratio $B/(B-L)$ has been determined in the standard model
considering chemical equilibrium and conservation
laws \cite{HarveyT}.

Recent low energy data \cite{Amaldi} strongly support supersymmetric
grand unified theories which have been proposed to solve the gauge
hierarchy problem \cite{Sakai}.
The effective supersymmetry breaking scale is expected to be around
the electroweak mass scale.
Therefore
we are led to consider
that supersymmetric particles may still be regarded as light (not decoupled)
at the lowest temperature $T_0$
for the electroweak reprocessing.
Then we need to reconsider the chemical equilibrium conditions
taking account of new particles (supersymmetric particles) and new
reactions (supersymmetric interactions).
Electroweak reprocessing in supersymmetric models has been discussed
by many groups in various contexts \cite{IbanezQ}, \cite{CohenN}.

The purpose of this paper is to study the chemical equilibrium
conditions in the electroweak baryon number reprocessing for
supersymmetric standard models (SSM) and to examine the threshold effect
of possible heavy particles, namely supersymmetric particles,
top quarks and Higgs particles.
Assuming a nonvanishing primordial $B-L$, we obtain a nonvanishing
$B$ and $L$ in supersymmetric standard models.
Since there are many more species of particles in supersymmetric
models than in nonsupersymmetric models, $B$ depends on the primordial
amount of supersymmetric particles and is not proportional to
$B-L$.
However, $B/(B-L)$ becomes a constant once soft supersymmetry
breaking is taken into account.
If the lowest effective temperature $T_0$ for the electroweak
reprocessing is higher than the electroweak phase transition
temperature $T_c$, we find a theorem: $B/(B-L)$ for the supersymmetric
models with $N_h$ pairs of Higgs doublets becomes identical to $B/(B-L)$
for the nonsupersymmetric models with $N_h$ Higgs doublets (half as
many Higgs as the supersymmetric case).
If $T_0 < T_c$, $B/(B-L)$ for the supersymmetric models turns out to
be about one percent less than that for the nonsupersymmetric models.
We also consider threshold effects for heavy particles, namely
supersymmetric particles, top quarks and Higgs particles,
in view of their unknown masses.
We find that the ratio $B/(B-L)$ is about five percent less than
the case of the standard model if the top
quark mass is much higher than $T_0$.
As a heavy particle effect, we also study an alternative mechanism
of baryon asymmetry operative even if primordial $B-L$ vanishes.
We find that a mass difference of scalar-leptons produces a
nonvanishing baryon asymmetry, provided that the lepton number for each
generation is separately conserved during the cosmological evolution.
We examine the most popular mechanism for the supersymmetry
breaking, namely the model with the hidden sector embedded in
supergravity \cite{ChamseddineAN}.
This model conserves the lepton number for each generation to a
sufficient accuracy.
Unfortunately mass differences among lepton generations turn out to be
too small.
Let us stress that we do not need the assumption of separate lepton
number conservation for each generation except for this last mechanism
due to scalar lepton mass differences.

After we have written up this work, we became aware of a recent paper
which contains results partially overlapping with ours
 and is complementary to our work \cite{DreinerR}.
They studied electroweak reprocessing in supersymmetric models
aiming at constraining the possible $B$ or $L$
violating interactions.

Supersymmetric standard models contain supermultiplets for
gauge bosons $g$, $W$, $B$ of
the $\rmSU(3)_\rmc\**\rmSU(2)_\rmL\**\rmU(1)_Y$
 and for $N_h$ pairs of Higgs doublets $\phi_1$ and $\phi_2$.
The remaining particle content is $N$ generations of quarks and
leptons and their superpartners.
The masses of up-type quarks (down-type quarks and leptons)
are given by the Yukawa coupling with $\phi_1$ ($\phi_2$).
The particle content of the supersymmetric standard models
is shown in Tab.\ \ref{tab:1}, where the ordinary particles (those present
in the nonsupersymmetric standard models) are listed
in the left column and their superpartners in the right column.

 First, we consider the fully supersymmetric case (neglecting
soft supersymmetry breaking) and neglect threshold effects
for  massive particles (all the particles are then assumed to be
ultra-relativistic).
We assign chemical potential $\mu_X$ for a particle $X$
and have altogether $21+6N$ chemical potentials as listed in
Tab.\ \ref{tab:1}.

When an interaction is rapid enough, it enforces an equilibrium
relation among the chemical potentials.
The $\rmW^\+-$ interactions with SU$(2)_\rmL$ doublet fields provide
the following relations between the chemical potentials within the
doublet:
\beq
 \m(I_3=-1/2)=\m(I_3=1/2)+\m_\rmW,
\label{eqn:7}
\eeq
where the $I_3=-1/2$ and $I_3=1/2$ doublet pairs
correspond to $(d_\rmL, u_\rmL)$, $(\tilde d_\rmL, \tilde u_\rmL)$,
$(e_{i\rmL}, \nu_i)$, $(\tilde{e}_{i\rmL}, \tilde{\nu}_i)$,
$(\phi_1^0, \phi_1^+)$, $(\tilde \phi_1^0, \tilde \phi_1^+)$,
$(\phi_2^-, \phi_2^0)$ and $(\tilde \phi_2^-, \tilde \phi_2^0)$,
respectively.
Neutral Higgs boson interactions with
quarks and charged leptons give
\beq
\m_{u_\rmR}=\m_{u_\rmL}+\m_{10},\quad
\m_{d_\rmR}=\m_{d_\rmL}+\m_{20},\quad
\m_{i\rmR}=\m_{i\rmL}+\m_{20}.
\eeq
Neutral gaugino interactions lead to the following
relations between the chemical potentials of the superpartners.
\bea
\m_{\widetilde\rmW^0}=\m_{\tilde g}=\m_{\tilde\rmB},
&&  \m_{\widetilde\rmW_\rmL}=\m_\rmW+\m_{\tilde\rmB},
  \quad \m_{\widetilde\rmW_\rmR}=\m_\rmW-\m_{\tilde\rmB},\nonumber \\
 \m_{\tilde u_\rmL}=\m_{u_\rmL}+\m_{\tilde\rmB},
  &&\m_{\tilde u_\rmR}=\m_{u_\rmR}-\m_{\tilde\rmB},
 \quad \m_{\tilde d_\rmL}=\m_{d_\rmL}+\m_{\tilde\rmB},
\quad \m_{\tilde d_\rmR}=\m_{d_\rmR}-\m_{\tilde\rmB},\nonumber \\
 \m_{\tilde i}=\m_i+\m_{\tilde\rmB},
 && \m_{\widetilde{i\rmL}}=\m_{i\rmL}+\m_{\tilde\rmB},
  \quad \m_{\widetilde{i\rmR}}=\m_{i\rmR}-\m_{\tilde\rmB}, \label{eqn:int}\\
 \m_{\widetilde{1+}}=\m_{1+}-\m_{\tilde\rmB},
  &&\m_{\widetilde{2-}}=\m_{2-}-\m_{\tilde\rmB},
 \quad \m_{\widetilde{10}}=\m_{10}-\m_{\tilde\rmB},
  \quad\m_{\widetilde{20}}=\m_{20}-\m_{\tilde\rmB}.\nonumber
\label{eqn:21}
\eea
Higgsino mass terms provide the mixing between $\tilde\f_1$ and
$-i\s_2\tilde\f_2^{\ c}$ implying
$\m_{\widetilde{20}}=-\m_{\widetilde{10}}$, %\ \
$\m_{\widetilde{2-}}=-\m_{\widetilde{1+}}$.
Therefore we obtain by using eq.\ (\ref{eqn:int})
\beq\m_{\tilde\rmB}=\frac12(\m_{10}+\m_{20}).\label{eqn:24}\eeq
No more new relations are obtained from other interactions in the
supersymmetric models.
The above relations allow us to express all the chemical potentials in
terms of $4+N$ independent chemical potentials, say,
$\m_\rmW$, $\m_{\tilde\rmB}=(\m_{10}+\m_{20})/2$,
$\m_0:=(\m_{10}-\m_{20})/2$, $\m_{u_\rmL}$ and $\m_i$.

\begin{figure}
 \leavevmode
 \epsfysize=8cm
 \centerline{\epsfbox{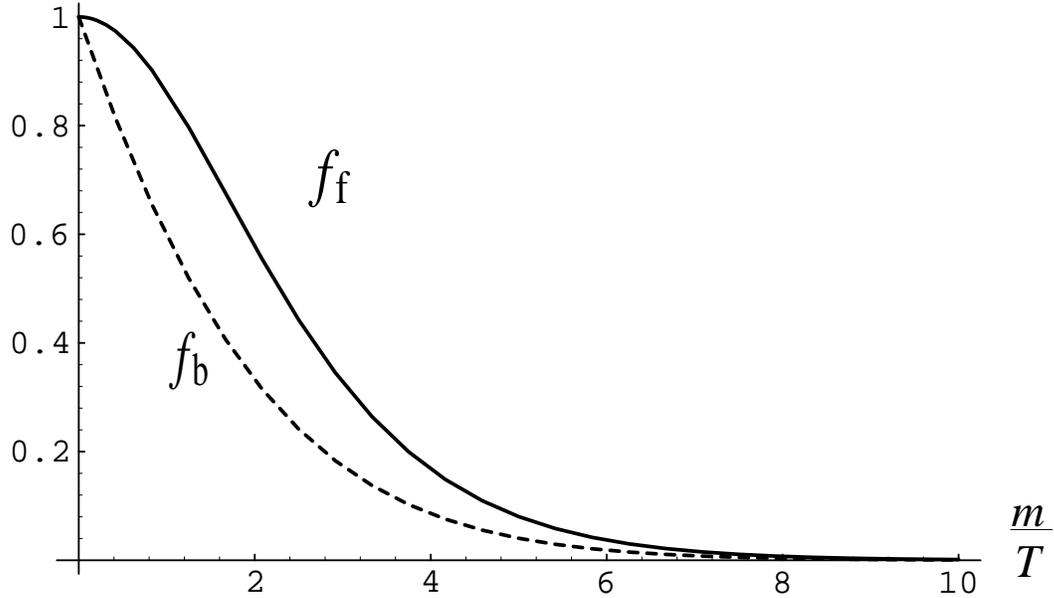}}
 \caption{Threshold functions for bosons $f_{\rm b}$ and fermions
 $f_{\rm f}$ as functions of the mass $m$ divided by the temperature
 $T$.  Solid line shows $f_{\rm f}$ and dashed line shows $f_{\rm b}$.}
 \label{fig:1}
\end{figure}

The difference of number density of particle $n_+$ and that of antiparticle
$n_-$ with mass $m$ at an equilibrium temperature $T$ is given for a
small chemical potential $\mu$
\bea
  n_+-n_- &\!\!\!=&\!\!\!
g\frac{T^3}{\p^2}\frac\m T
           \int_0^\infty\!dx\,
           \frac{x^2\exp\left\{-\rt{x^2+(\frac mT)^2}\right\}}
        {\left[\th+\exp\left\{-\rt{x^2+(\frac mT)^2}\right\}\right]^2}
\nonumber \\
&\!\!\!=:&\!\!\!
f_{\rm b, f}(m/T) \times (n_+-n_-)|_{m=0},
\label{particleasymmetry}
\eea
where
 $g$ is the number of
internal degrees of freedom
and $\th = -1$ for bosons and $\th = +1$ for fermions.
The particle asymmetry becomes in the ultra-relativistic limit $m << T$
\beq
  (n_+-n_-)|_{m=0}
=
\left\{
\begin{array}{cc}
\frac{gT^3}3\((\frac\m T\)) & {\rm (massless\ boson)}, \\
\frac{gT^3}6\((\frac\m T\)) & {\rm (massless\ fermion)}. \\
\end{array}
\right.
 \label{eqn:1}\\
\eeq
Let us call the ratio $f_{\rm b} (f_{\rm f})$
between the particle asymmetry of the massive
case and that of the massless case as
threshold functions for bosons (fermions).
They are functions of $m/T$ with $f_{\rm b,f}(0)=1$ and
$f_{\rm b,f}(\infty)=0$ as shown in Fig.\ \ref{fig:1}.

If we are interested in the total lepton number, but not in lepton
numbers of each generation separately, we need only the sum of lepton
chemical potentials $\mu=\sum_i \mu_i$ to express baryon, lepton and
charge densities.
Using eqs.\ (\ref{eqn:1}) and (\ref{eqn:7})--(\ref{eqn:24}),
\footnote{
Our equations eqs.\ (\ref{eqn:26})--(\ref{eq:charge}) disagree slightly
from those in \cite{IbanezQ}.
Even if we discard the Higgsino mass term, our formulas do not reduce
to theirs.
}
\bea
 B&\!\!\!=&\!\!\!\frac{T^2}6\left[3\cdot2N(2\m_{u_L}
+\m_\rmW+\m_{\tilde\rmB})\right],\label{eqn:26}\\
 L&\!\!\!=&\!\!\!\frac{T^2}6\left[3(3\m+2N\m_\rmW-N\m_0)
+5N\m_{\tilde\rmB}\right],\\
 Q&\!\!\!=&\!\!\!\frac{T^2}6\left[3\cdot2\{N\m_{u_L}-\m
-(2N+1+N_h)\m_\rmW+(2N+N_h)\m_0\}\right].\label{eq:charge}
\eea
When a Higgsino mass term is present, there are no other global
continuous symmetries than $B$ and $L_i$.
Neglecting the mixed gravitational anomaly \cite{IbanezQ},
we only have the electroweak anomaly, which implies
$ N(\m_{u_\rmL}+2\m_{d_\rmL})+\sum_i \m_i=0,$
or, from eq.\ (\ref{eqn:7}),
\beq3N\m_{u_\rmL}+2N\m_\rmW+\m=0.\label{eqn:32}\eeq

We require $Q=0$ in our universe.
Above the electroweak phase transition temperature $T_c$,
the unbroken $\rmSU(2)_\rmL$
symmetry
implies $\m_\rmW=0$. Then we find
\beq
 B
=
\frac{T^2}6 \left[3\cdot2N(2\m_{u_\rmL}
+\m_{\tilde\rmB})\right],\quad
 B-L
=
\frac{T^2}6 \left[3N\frac {22N+13N_h}{2N+N_h}\m_{u_\rmL}
+N\m_{\tilde\rmB}\right].
\eeq
Let us note that $B-L=0$ implies $B+L\propto\m_{\tilde\rmB}\neq0$ in
general. Namely we can have nonvanishing $B+L$ due to
the primordial supersymmetric particles even when the primordial
$B-L$ vanishes. This situation in the fully supersymmetric models
is quite different from the standard model.

We now incorporate
soft supersymmetry breaking.
The following soft supersymmetry breaking terms provide
nontrivial relations among chemical potentials:
\ben
 \item Higgs mass mixing terms:
 \beq
  B(i\s_2\f_1)^T\f_2+\rmHc,
 \eeq
 \item gaugino mass terms:
 \beq
  -\frac12 m_{\widetilde\rmW}(\bar{\widetilde\rmW}^+{\widetilde\rmW^+}
+\bar{\widetilde\rmW}^-{\widetilde\rmW^-}+\bar{\widetilde\rmW}^0{\widetilde\rmW^0})
  -\frac12 m_{\tilde\rmB}\bar{\tilde\rmB}\tilde\rmB
  -\frac12 m_{\tilde{\rm g}}\bar{\tilde g}\tilde g,
 \eeq
 \item scalar tri-linear terms:
 \beq
  -A_u\tilde u_\rmR^*(i\s_2\f_1)^T\tilde{q}_\rmL-A_d\tilde d_\rmR^*
 (-i\s_2\f_2)^T\tilde{q}_\rmL-A_e\tilde e_\rmR^*(-i\s_2\f_2)^T
 \tilde{l}_\rmL+\rmHc
 \label{trilinear}
 \eeq
\een
They imply the following relations in addition to
those in the fully supersymmetric case:
\beq
 \m_{\tilde\rmB}=\frac12 (\m_{10}+\m_{20})=0,\quad
 \m_{10}=-\m_{20}=\m_0.
\eeq
Although we have suppressed the generation indices in
eq.~(\ref{trilinear}), the soft supersymmetry breaking can give an
arbitrary pattern of generation mixing in general.
There are also quadratic terms which mix among scalar particles.
Therefore we do not assume separate lepton number conservation for
each generation except at the end of this work.
Since the soft supersymmetry breaking terms should be present,
we take these relations into account from now on.
Thus we have altogether $3+N$ independent chemical potentials.

Gauge interactions, Yukawa interactions and their supersymmetric
extensions are in thermal equilibrium well below $T_c$.
However, we should take account of threshold effects of heavy
particles,
since we expect the supersymmetric particles, top and Higgs particles
to be heavy.

Now, using eq.\ (\ref{particleasymmetry}), we can examine the
threshold effect of heavy particles.
For simplicity, we approximate all the standard
model particles to be massless, all the scalar-quarks and scalar-leptons
to have a common mass $m_{\rm Sb}$ and all the charginos to have
 a common mass $m_{\rm Sf}$. Let us define
\beq f_{\rm Sb}=f_{\rm b}\((\frac{m_{\rm Sb}}{T_0}\)), \qquad
 f_{\rm Sf}=f_{\rm f}\((\frac{m_{\rm Sf}}{T_0}\)),\label{eqn:40}\eeq
as threshold functions defined in eq.\ (\ref{particleasymmetry}) for
supersymmetric bosons and fermions,
respectively.
Using eqs.\ (\ref{eqn:7})--(\ref{eqn:24}) and (\ref{eqn:40}), we find
\bea
 B&\!\!\!=&\!\!\!\frac{{T_0}^2}6\left[N\left\{\m_{u_\rmL}+\m_{u_\rmR}
+\m_{d_\rmL}+\m_{d_\rmR}+2f_{\rm Sb}\((\m_{\tilde u_\rmL}
+\m_{\tilde u_\rmR}+\m_{\tilde d_\rmL}
+\m_{\tilde d_\rmR}\))\right\}\right] \nonumber \\
  &\!\!\!=&\!\!\!
\frac{{T_0}^2}6 2(1+2f_{\rm Sb})N(2\m_{u_\rmL}+\m_\rmW),
  \\
  L&\!\!\!=&\!\!\!\frac{{T_0}^2}6 \left[\sum_i(\m_i+\m_{i\rmL}+\m_{i\rmR})
+2f_{\rm Sb}\sum_i\((\m_{\tilde i}+\m_{\widetilde{i\rmL}}
+\m_{\widetilde{i\rmR}}\))\right] \nonumber \\
   &\!\!\!=&\!\!\!\frac{{T_0}^2}6 (1+2f_{\rm Sb})(3\m+2N\m_\rmW-N\m_0),
\label{leptonnumber}
  \\
  Q
&\!\!\!=&\!\!\!
\frac{{T_0}^2}6 \left[2N(\m_{u_\rmL}+\m_{u_\rmR})
-N(\m_{d_\rmL}+\m_{d_\rmR})
-\sum_i(\m_{i\rmL}+\m_{i\rmR})
\right.\nonumber\\
&\!\!\!&\!\!\!
-4\m_\rmW+2N_h(\m_{1+}-\m_{2-})
+f_{\rm Sf}\left\{-\m_{\widetilde\rmW_\rmL}
-\m_{\widetilde\rmW_\rmR}+N_h\((\m_{\widetilde{1+}}
-\m_{\widetilde{2-}}\))\right\}
\nonumber\\
&\!\!\!&\!\!\!
\left.+f_{\rm Sb}\left\{4N\((\m_{\tilde u_\rmL}
+\m_{\tilde u_\rmR}\))-2N\((\m_{\tilde d_\rmL}
+\m_{\tilde d_\rmR}\))-2\sum_i\((\m_{\widetilde{i\rmL}}
+\m_{\widetilde{i\rmR}}\))\right\}\right],\\
   &\!\!\!=&\!\!\!\frac{{T_0}^2}6 [2(1+2f_{\rm Sb})(N\m_{u_\rmL}-\m)
-2\left\{2(1+2f_{\rm Sb})N+(2+f_{\rm Sf})(1+N_h)\right\}
\m_\rmW\nonumber\\
&\!\!\!&\!\!\!
+2\left\{2(1+2f_{\rm Sb})N
+(2+f_{\rm Sf})N_h\right\}\m_0].
\eea
We have thus obtained a formula which interpolates between the
nonsupersymmetric
standard models (at the large mass limit for supersymmetric
particles $f_{\rm Sb,f} \rightarrow 0$)
and the fully supersymmetric models (at the small mass limit
 $f_{\rm Sb,f} \rightarrow 1$).
One should note that we have only four independent combinations of
chemical potentials to describe the above quantities, since
we are only interested in total lepton number rather than
lepton number for each generation.

Let us consider the case where the electroweak reprocessing ceases
to be effective above the critical temperature $T_\rmc$ for the
electroweak phase transition.
Above the critical temperature $T_0>T_\rmc$, we should impose
$\m_\rmW=0$ and $Q=0$.
Taking account of the relation (\ref{eqn:32}) due to the soft
supersymmetry breaking terms, we can express $B$
and $L$ in terms of a single chemical potential as in
the standard model case:
\beq
  B
=
\frac{{T_0}^2}6 4(1+2f_{\rm Sb})N\m_{u_\rmL},
\quad  B-L
=
\frac{{T_0}^2}6 (1+2f_{\rm Sb})N\frac{22N
+13SN_h}{2N+SN_h}\m_{u_\rmL},
\eeq
where we use the following ratio that is convenient to express
the effect of supersymmetric particles:
\beq
S=\frac{2+f_{\rm Sf}}{1+2f_{\rm Sb}}.
\label{coeffnh}
\eeq

 From these, we obtain the following relation:
\beq
  B
=
\frac{8N+4 SN_h}
      {22N+13 SN_h}(B-L).
\eeq
The fully supersymmetric case is obtained by
$f_{\rm Sb,f}\rightarrow 1$.
On the other hand, the nonsupersymmetric model case results
if supersymmetric particles are heavy $f_{\rm Sb,f}\rightarrow 0$.
It is interesting to observe that these two limits are
different only by the coefficient $S$ in front of $N_h$.
The $S$ for the supersymmetric case becomes half of the $S$ for
the nonsupersymmetric case.
The physical reason behind this fact is the following.
The coefficient in front of $N$ is essentially counting
the degrees of freedom for quarks and leptons, whereas that in front of
$N_h$ is essentially counting the degree of freedom of Higgs particles.
As shown in eq.\ (\ref{eqn:1}), bosons contribute
for this coefficient twice as much as fermions.
In the supersymmetric case, we have superpartners of quarks and
leptons which are bosons, and superpartners of Higgs particles
which are fermions.
Therefore the scalar-quarks and scalar-leptons multiply
the coefficient in front of $N$ by $1+2=3$, whereas Higgsinos
multiply
the coefficient in front of $N_h$ by $1+1/2=3/2$.
Therefore the relative importance of Higgs-Higgsino
contributions to $B/(B-L)$ in the supersymmetric case
becomes half as much as the Higgs contributions
in the nonsupersymmetric case.
As a corollary, we find the following theorem:
the supersymmetric standard model
with $N_h$ pairs of Higgs-Higgsino supermultiplets gives
identical $B/(B-L)$ to the standard model with only $N_h$
Higgs particles (half as many Higgs particles compared to
supersymmetric case) above the electroweak phase transition temperature.

Let us now turn to the case where the lowest effective temperature
for the electroweak reprocessing is below
the critical temperature $T_\rmc$ for the
electroweak phase transition.
Below the critical temperature $T_0<T_\rmc$, $Q=0$ but $\mu_W\not=0$.
On the other hand, $\m_0$ must vanish because of the vacuum condensation
of neutral Higgs bosons. By using eq.\ (\ref{eqn:32}), we can again express
$B$ and $L$ in terms of a single chemical potential and we obtain
the following relation using $S$ defined in eq.\ (\ref{coeffnh}):
\beq
   B=\frac{8N+4 S(N_h+1)}
       {24N+13 S(N_h+1)}(B-L).
\eeq
For the minimal supersymmetric standard model with three
generations ($N=3)$ and one pair of Higgs doublet ($N_h=1$)
by setting $f_{\rm Sb,f}=1$, we obtain
\beq B=\frac{16}{49}(B-L)=0.3265\cdot\cdot\cdot\times(B-L).\eeq
For the minimal standard model with three generations and one
Higgs doublet \cite{HarveyT},
\beq B=\frac{12}{37}(B-L)=0.3243\cdot\cdot\cdot\times(B-L).
\label{eqn:59}\eeq
Supersymmetry gives a correction of about one percent for $B/(B-L)$.

The top-quark mass $m_t$ or the charged Higgs mass $m_{\phi}$
 may be large or comparable to the lowest effective
temperature $T_0$ for the electroweak anomaly interaction.
Therefore we should also consider the
threshold effect of a heavy top-quark.
Let us take
the number of generations to be three ($N=3$) and the number of
pairs of Higgs doublets to be one ($N_h=1$).
The neutral physical Higgs particle does not contribute to $B$, $L$, $Q$.
However, charged Higgs particles do contribute and they may be heavy
too.
Therefore we introduce the threshold functions for the top $f_\rmt$
and charged Higgs particles $f_\phi$:
\beq
 f_\rmt:= f_{\rm f}\((\frac{m_\rmt}{T_0}\)),\qquad
 f_\f:= f_{\rm b}\((\frac{m_\f}{T_0}\)).
\eeq
Imposing $\m_0=Q=0$ and using eq.\ (\ref{eqn:32}), we obtain
\bea
&\!\!\!&\!\!\!
\quad  {B \over B-L}= \label{eqn:bblft}\\
&\!\!\!&\!\!\!
\frac{2(5+12f_{\rm Sb}+f_\rmt)(9+12f_{\rm Sb}+2f_{\rm Sf}+f_\f)}
        {291+6(171+141f_{\rm Sb}+26f_{\rm Sf}+12f_\rmt+13f_\f)f_{\rm Sb}
       +2(37+f_\rmt)f_{\rm Sf}+2(21+f_\f)f_\rmt+37f_\f}.  \nonumber
\eea
The most significant threshold effect is obtained when the top
quark mass is important.
Supposing $f_{\rm Sb,f}=f_\f=0$ in particular, we obtain the standard
model case where only the top-quark mass may not be negligible:
\beq
 B=\frac{6(5+f_\rmt)}{97+14f_\rmt}(B-L)
\rightarrow
\frac{30}{97} (B-L)=0.309\cdot\cdot\cdot\times(B-L),
\eeq
where the top quark mass is much larger than the lowest effective
temperature $T_0$ for the electroweak reprocessing.
We see that a heavy top quark gives a $B/(B-L)$ that is about five
percent
less than that for the light top quark case, given in eq.\ (\ref{eqn:59}).
We can see also the effect of the charged Higgs from eq.\ (\ref{eqn:bblft}).
If the standard model with one Higgs doublet is modified by adding an extra
Higgs doublet ($f_{\rm Sb,f}=0$, $f_\f=f_\rmt=1$ in eq.\ (\ref{eqn:bblft})),
the $B/(B-L)$ decreases about 0.5 \%.

It has been noted \cite{KuzminRS}, \cite{DreinerR} that the
nonperturbative electroweak reprocessing
of baryon and lepton numbers actually gives conserved quantities
 $l_i=L_i-B/N$ for each generation separately as implied by
eq.\ (\ref{eqn:32}),
where $L_i$ is the lepton number of the $i$-th generation.
This conservation law offers a possibility that a baryon asymmetry
may be generated even if the primordial $B-L$ vanishes,
provided
masses and the primordial $l_i$ differ for leptons of different
generations.
We find that this mechanism can be more significant for
scalar-leptons, since bosons have stronger
 temperature dependence than fermions
($f_{\rm b}\approx1-\frac32\frac m{\p T}$,
$f_{\rm f}\approx1-\frac32\((\frac m{\p T}\))^2$ ) for $m\ll T$ as
shown in Fig.\ \ref{fig:1}.

{}From now on, we assume the separate conservation of $l_i$ during the
cosmological evolution and examine the consequences of the electroweak
transmutation of $l_i$ into $B$.
It is convenient to use the $N$ conserved quantities $l_i$
instead of the $N$ chemical potentials $\mu_i$.
Three other chemical potentials $\mu_W$, $\mu_0$, $\mu_{uL}$,
which remain from the $3+N$ chemical potentials,
can be determined by $Q=0$, $\mu_W=0$ ($\mu_0=0$)
for $T_0>T_c$ ($T_0<T_c$) and by the electroweak selection rule
eq.\ (\ref{eqn:32}).
Let us denote the threshold function of a scalar-lepton of $i$-th
generation as $f_i$.
For simplicity, we
assume
$l_N\not=0$, $L-B \equiv \sum l_i=0$.
As a typical example,
we study the mechanism by assuming that the scalar-tau-lepton is heavier than
the scalar-quarks and the rest of scalar-leptons:
$f_1=\cdots=f_{N-1}=f_{\rm Sb}$.

\begin{figure}
 \leavevmode
 \epsfysize=8cm
 \centerline{\epsfbox{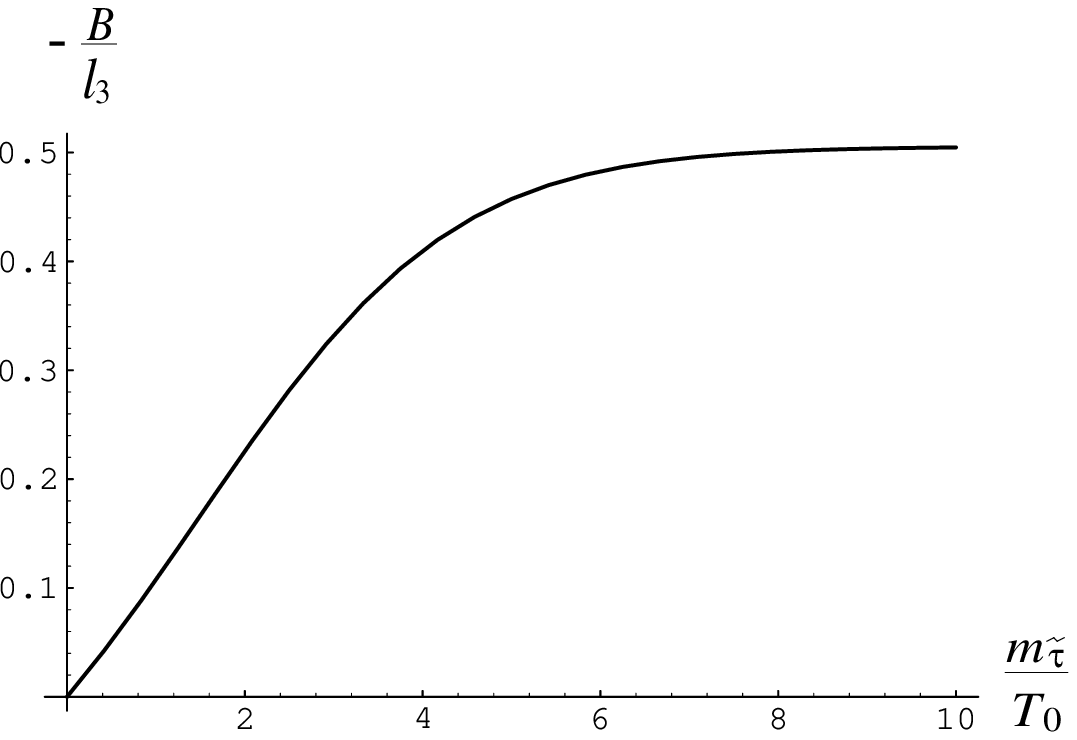}}
 \caption{The ratio of the baryon number to the primordial lepton number
 $-B/l_3$ as a function of $m_{\tilde \ta}/T_0$ for
 $T_0 > T_c$.}
 \label{fig:2}
\end{figure}

In the case of $T_0 > T_c$, we
obtain the baryon
number in terms of $f_{\rm Sb}$ and $f_N$
using $S$ defined in eq.\ (\ref{coeffnh}):
\beq
-{B \over l_N}={{\frac83 \al\left\{5N+3SN_h
+\frac{2(f_{\rm Sb}-f_N)}{1+2f_{\rm Sb}} \right\}} \over
{22N+13SN_h+\frac2N \al\left\{11N+4SN_h-\frac23(13N-4)
\frac{f_{\rm Sb}-f_N}{1+2f_{\rm Sb}}
\right\}}},
\eeq
where we define $\al=\frac{f_{\rm Sb}-f_N}{1+2f_N}$.
For an illustration, we
assume $N=3$, $N_h=1$ and that supersymmetric particles are light except the
scalar-tau-lepton: $f_{\rm Sb}=f_{\rm Sf}=1$.
The ratio
$-B/l_3$
is shown in Fig.\ \ref{fig:2}
as a function of the mass of the
scalar-tau-lepton
 divided by the temperature $m_{\tilde \ta}/T_0$.

\begin{figure}
 \leavevmode
 \epsfysize=8cm
 \centerline{\epsfbox{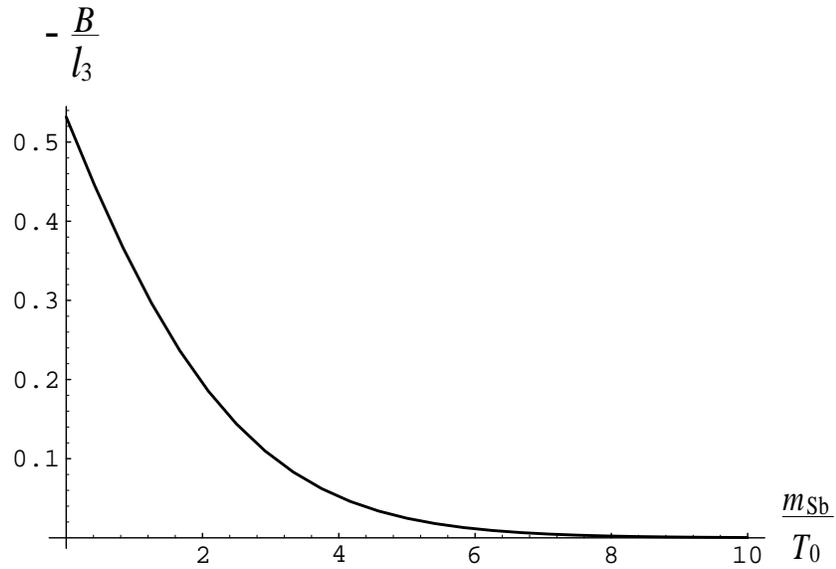}}
 \caption{The ratio of the baryon number to the primordial lepton number
 $-B/l_3$ as a function of $m_{\rm Sb}/T_0$ for
 $T_0 < T_c$.}
 \label{fig:3}
\end{figure}

In the case of $T_0 < T_c$, we obtain the baryon
number in terms of $f_{\rm Sb}$ and $f_N$:
\beq
-{B \over l_N}=\frac{\frac43\al\left\{11N+6SN_h(N_h+1)+
\frac{8(f_{\rm Sb}-f_N)}{1+2f_{\rm Sb}}\right\}}
{24N+13S(N_h+1)+\frac8N\al\left\{
4N+S(N_h+1)-\frac13(13N-4)\frac{f_{\rm Sb}-f_N}{1+2f_{\rm Sb}}
\right\}}.
\eeq
We
assume $N=3$, $N_h=1$ and that the scalar-tau-lepton and charginos are heavy
($f_3=f_{\rm Sf}=0$).
In Fig.\ \ref{fig:3},
$-B/l_3$ is shown as a function of the mass of the
other scalar supersymmetric particles
divided by the temperature $m_{\rm Sb}/T_0$.
We see that this mechanism is very efficient: it can convert about
half of the primordial $l_3$ to $B$.

In general we can obtain sufficient mass differences among scalar
leptons, since the soft supersymmetry breaking allows arbitrary
differences of the order of the electroweak mass scale.
However, the separate conservation of $l_i$ during the cosmological
evolution imposes a stringent limit on the possible mixing among
generations in order for our mechanism to be viable.
To estimate the limit for mixing angles $\th$, we take a typical
generation changing interaction:
\beq
\tilde{\nu}_\tau+ W \rightarrow \tilde{\nu}_\mu + W,
\label{eq32}
\eeq
where the scalar-tau-neutrino interacts with $W$ boson and is
converted into the scalar-mu-neutrino.
The convertion rate $\Gamma$ is given by
\beq
\G \cong n_W \s v
\sim  \frac{(\th g^2)^2 T^5}{(T^2+M^2)^2}
\sim \th^2 g^4 T,
\eeq
where $n_W$ is the number density of $W$, $\s$ is the cross
section for (\ref{eq32}),
$v$ is the velocity of scalar-tau-neutrino,
$\th$ is the mixing angle between scalar-tau-neutrino and scalar-mu-neutrino,
and $M$ is the mass of the scalar-lepton.
The primordial $l_i$ may be maintained during the cosmological
evolution, if the conversion rate $\Gamma$ is much less than the
expansion rate
$H := {1 \over R}{d R \over d t} \cong g_*^{1/2} T^2/m_{\Pl}
\sim 10^{-18} (T/\mbox{GeV})^2$ where $g_* \sim 10^2$ is the total
effective number of degrees of freedom.
Using $g^2 \sim 10^{-1}$ we find
\beq
\th \ll 10^{-8} (T/\mbox{GeV})^{1/2}.
\label{mixingbound}
\eeq
This bound is quite stringent. For instance,
$\th \ll 10^{-6}$ ($10^{-7}$) for $T\sim$ 10 TeV (100 GeV).

We have studied the supergravity model as the most popular model for
the supersymmetry breaking.
The spontaneous breaking occurs only in the hidden sector, and is
transmitted by the supergravity to the observable sector
\cite{ChamseddineAN}.
We find that the mixing among lepton generations is
small enough to satisfy (\ref{mixingbound}).
However, the mass differences among scalar leptons of diferent
generations turn out to be small even if the seesaw mechanism due to
large Majorana mass for right-handed neutrino is introduced.
Therefore we see that we need to consider another model for
supersymmetry breaking in order for our mechanism using separate
$l_i$ conservation to be effective.

We would like to thank H. Murayama for a useful discussion, and
Ken-ichiro Aoki and Martin Hirsch for a careful reading of the manuscript.
This work is supported in part by Grant-in-Aid for Scientific
Research (T. Inui) and (No.\ 05640334) (N. S.), and Grant-in-Aid for
Scientific
Research for Priority Areas (No.\ 05230019) (N. S.) {}from
the Ministry of Education, Science and Culture.
\par

\vspace{3mm}
\begin{table}[p]
 \begin{tabular}{|c|c||c|c|}
  \hline
  \ SM particles\ &chemical potentials
   &superpartners&chemical potentials\\
  \hline
  $\ba cu_i\\d_i\ea_\rmL$&$\begin{array}c\m_{u_\rmL}\\\m_{d_\rmL}\end{array}$
   &$\ba c\tilde u_i\\ \tilde d_i\ea_\rmL$&$\begin{array}c\m_{\tilde
u_\rmL}\\\m_{\tilde d_\rmL}\end{array}$\\
  \hline
  $\ba c\nu_i\\e_i\ea_\rmL$&$\begin{array}c\m_i\\ \m_{i\rmL}\end{array}$
   &$\ba c\tilde{\nu}_i\\\tilde e_i\ea_\rmL$&$\begin{array}c \m_{\tilde
i}\\\m_{\widetilde{i\rmL}}\end{array}$\\
  \hline
  $u_{i\rmR}$&$\m_{u_\rmR}$ & $\tilde u_{i_\rmR}$&$\m_{\tilde u_\rmR}$ \\
  \hline
  $d_{i\rmR}$ & $\m_{d_\rmR}$ & $\tilde d_{i_\rmR}$ &$\m_{\tilde d_\rmR}$ \\
  \hline
  $e_{i\rmR}$ & $\m_{i\rmR}$ & $\tilde e_{i\rmR}$ & $\m_{\widetilde{i\rmR}}$ \\
  \hline
  $\rmW^-$ & $\m_\rmW$ & $\widetilde\rmW^-_\rmL$ &
$\m_{\widetilde\rmW_\rmL},\ \m_{\widetilde\rmW_\rmR}$ \\
  \hline
  $\rmW^0$ & $\m_{\rmW^0}=0$ & $\widetilde\rmW^0_\rmL$ &
$\m_{\widetilde\rmW^0}$ \\
  \hline
  $\rmB^0$ & $\m_\rmB=0$ & $\tilde{\rmB}^0_\rmL$ & $\m_{\tilde\rmB}$ \\
  \hline
  $g$ & $\m_g=0$ & $\tilde{g}_\rmL$ & $\m_{\tilde g}$ \\
  \hline
  $\ba c\f_{1i}^+\\\f_{1i}^0\ea$&$\begin{array}c\m_{1+}\\\m_{10}\end{array}$
   &$\ba
c\tilde\f_{1i}^+\\\tilde\f_{1i}^0\ea_\rmL$&$\begin{array}c\m_{\widetilde{1+}}\\\m_{\widetilde{10}}\end{array}$\\
  \hline
  $\ba c\f_{2i}^0\\\f_{2i}^-\ea$&$\begin{array}c\m_{20}\\\m_{2-}\end{array}$
   &$\ba
c\tilde{\f}_{2i}^0\\\tilde{\f}_{2i}^-\ea_\rmL$&$\begin{array}c\m_{\widetilde{20}}\\\m_{\widetilde{2-}}\end{array}$\\
 \hline
 \end{tabular}
 \caption{Particles and chemical potentials of SSM.}
 \label{tab:1}
\end{table}
%
%%%%%%%%%%%%%%%%%%%
%%%%%%%%%%%%%%%%%%%

\vspace{3mm}
\end{document}